\documentclass[12pt]{article}
\usepackage{graphicx}
\usepackage{dcolumn}
\usepackage{bm}

\begin{document}

\begin{flushright}
UK/07-06
\end{flushright}

\begin{center}

{\bf {\Large Neutron Electric Dipole Moment at Fixed Topology}}

\vspace{0.6cm}


{\bf  K.F. Liu}

\end{center}

\begin{center}
\begin{flushleft}
{\it
Dept.\ of Physics and Astronomy, University of Kentucky, Lexington, KY 40506
}
\end{flushleft}
\end{center}

\begin{abstract}
 We describe the finite volume effects of CP-odd quantities, such as the
neutron electric dipole moment and the anapole moment in the $\theta$-vacuum,
under different topological sectors. We evaluate the three-point Green's functions for
the electromagnetic current in a fixed non-trivial topological sector in order to
extract these CP-odd observables.  We discuss the role of zero modes in the CP-odd
Green's function and show that, in the quenched approximation, there is a power divergence
in the quark mass for CP-odd quantities at finite volume.
\end{abstract}

\section{Introduction}

    CP-violation is still not well understood. The baryon asymmetry in the universe
suggests that there is another source of CP-violation besides the CP phase in the
CKM matrix as measured from the K and B meson decays.. The recent discovery of non-zero
neutrino mass has prompted suggestions of CP-violation in the leptonic sector.
On the other hand, it is known for many years that QCD has a $\theta$ term which
gives rise to strong CP-violation. However, its consequence in quantities such as the
neutron electric dipole moment has not been discovered. This puts the limit of
$\theta$ to be less than $10^{-9}$. Next generation of experiments will further
push down the limit and hopefully will detect it at some stage.
The relation between the neutron electric dipole moment due to the $\theta$ term can
be calculated via lattice QCD and there are several attempts to do so~\cite{bbo06,sai05,sai07,sai08}.
Since $\theta$ in QCD is small, one can compute the CP-odd part of the
Green's function as the first derivative w.r.t. $\theta$, i.e.
$i\theta \langle Q \mathcal{O}\rangle_{\theta=0}= i\theta \sum_{\nu} P_{\nu}\nu
\langle \mathcal{O}\rangle_{\nu}$. This requires a weighted sum of
the Green's function $\langle \mathcal{O}\rangle_{\nu}$ in different topological
sectors. In view of the fact that practical Monte Carlo simulations can get
stuck in a particular topological sector, how is one to evaluate CP-odd quantities such
as the neutron electric dipole moment and anapole moment in this case? The situation
of being stuck in one topological sector is getting more serious as the gauge field
becomes smooth,
the quark mass is approaching the physically
small u/d mass and the continuum limit is approached. Since simulating overlap
fermion with HMC faces the difficulty of tackling discontinuity of the sign function
associated with topological change, it has prompted  JLQCD to add a extra fermion action
to prevent the topological change to avoid this numerical difficulty~\cite{fhh06}.
Recently it is shown that gauge action can be obtained from the overlap operator
as $Tr D_{ov}$~\cite{iva06,ahl08} and can be simulated together with the overlap fermion~\cite{liu06}.
The first attempt in implementing it in HMC with the rational approximation has
revealed that even for heavy quark ($ma=0.5$) for a range of lattice spacing
$a=0.06 - 0.15$ fm, no topological change has been encountered~\cite{streuer07}.
These initial MC runs have raised the issue of ergodicity.
Nevertheless, if the Monte Carlo algorithm is ergodic, all the physical quantities
including the CP-odd quantities can be obtained within a fixed topological sector
provided finite volume correction is taken into account. This has been formulated
and studied for CP-even quantities such as the mass~\cite{bcn03} and it has been recently
extended to CP-odd quantities and shown that topological susceptibility can be obtained in a fixed
topological sector~\cite{afh07}. In the present work, we shall concentrate on the
three-point function and discuss the electromagnetic form factors and the
neutron electric dipole form factors in the fixed topological sector. In the above
expression to extract NEDM, the Green's function $\langle \mathcal{O}\rangle_{\nu}$ is
weighted with $\nu$. Since the average of topological charge $|\nu|$ grows with $\sqrt{V}$,
in what way does the above expression converge at large volume? We shall address this
point. We shall point out that zero modes are part of the physical spectrum and they
contribute to the leading $1/V$ behavior in the CP-odd case. As such, the
quenched calculation of NEDM and, for that matter, other CP-odd matrix elements
suffer power divergence in quark mass at the chiral limit. Finally, we will discuss the possible
relation between the anomalous magnetic form factor and NEDM of the nucleon.

\section{$\theta$ Vacuum and Topological Sectors}

    We shall start with a brief review of the relation between the $\theta$
vacuum and the associated topological sectors in QCD.
     The QCD partition function in a $\theta$ vacuum on a torus in the continuum formulation
is written as

\begin{equation}    \label{Z-theta}
Z(\theta)=\int \mathcal{D}A\mathcal{D}\overline{\Psi}\mathcal{D}\Psi e^{-S[A,
\overline{\Psi},\Psi]} e^{i\theta Q[A]},
\end{equation}
where the topological charge operator $Q[A]$ is defined as
\begin{equation}
Q[A] = \frac{1}{16\pi^2}\int d^4x Tr(G_{\mu\nu}\tilde{G}_{\mu\nu}).
\end{equation}

    For a torus of $V= L^3 L_t$ with periodic or anti-periodic boundary condition,
the topological charge operator defined with Gisparg-Wilson fermion, such as the
overlap fermion~\cite{neu98} will have integer charges $\nu$ eigenvalues on smooth gauge
configurations. Since the gauge configurations occurring in the path-integral
in Eq.~(\ref{Z-theta}) are characterized by an integer, the partition function of
the $\theta$ vacuum is a sum of partition functions characterized by this integral
topological charge $\nu$
\begin{equation}
Z(\theta)= \sum_{\nu = - \infty}^{\nu=\infty} e^{i\theta \nu} Z_{\nu}
\end{equation}
With $\nu$ being an integer, $Z(\theta)$ is periodic, i.e. $Z(\theta + 2\pi) = Z(\theta)$,
and therefore $Z_{\nu}$ is the Fourier transform of $Z(\theta)$ in the range $\theta
\in [-\pi, \pi]$
\begin{equation}   \label{Z-nu}
Z_{\nu}= \frac{1}{2\pi}\int_{-\pi}^{\pi} d\theta Z(\theta)\,e^{-i \theta \nu}
\end{equation}

    The partition function $Z(\theta)$ in Eq.~(\ref{Z-theta})
can be expressed in terms of the energy eigenstates
\begin{equation}
Z(\theta)= \sum_{n} e^{-V e_n(\theta)},
\end{equation}
where $e_n(\theta)$ is the energy density of the nth eigenstate. When the spatial volume
is large, the energy of the $\theta$-vacuum is $E_0= V e_0(\theta)$. For $\theta= \pi$,
the derivative of $Z(\theta)$ is discontinuous at $\theta = \pm \pi$
and CP symmetry is spontaneously broken~\cite{wit80}. For small $\theta$, we assume
that $Z(\theta)$ is analytic and, on account of the fact that $Z(\theta)= Z(-\theta)$,
$e_0(\theta)$ can be expanded around $\theta=0$
\begin{equation}  \label{theta-expansion}
e_0(\theta) = \frac{1}{2}\chi_t \theta^2 + \gamma \theta^4+ ...
\end{equation}
where
\begin{equation}
\chi_t = \frac{\langle Q^2\rangle}{V},
\end{equation}
is the topological susceptibility. At low temperature with $L_t (E_1(\theta) - E_0(\theta))
\gg 1$ ($L_t=1/T$), the vacuum state dominate and, therefor,
\begin{equation}   \label{vacuum-partition}
Z(\theta)= e^{-Ve_0(\theta)}=e^{-\frac{V\chi_t}{2} \theta^2} (1 - \gamma \theta^4 +
\mathcal{O}(\theta^6)).
\end{equation}

   When the temperature is not low compared to the energy gap which is the pion mass, the
partition function is the Boltzmann factor which contains contributions from the excited
states depending on the temperature. In this case, Eqs.~(\ref{vacuum-partition}) and
(\ref{theta-expansion}) still hold provided that the three-volume
is large and the vacuum energy density is replaced by the free energy
density $F(T,\theta)$~\cite{ly52,ls92}. We shall only consider the low temperature situation
in this work.

   At low temperature and small topological charge with $|\nu|/\sqrt{V}$ fixed, the
partition function for the topological sector $\nu$ in Eq.~(\ref{Z-nu}) becomes
\begin{equation}
Z_{\nu}= \frac{1}{2\pi}\int_{-\pi}^{\pi} d\theta e^{-Ve_0(\theta)}\,e^{-i \theta \nu}.
\end{equation}
where the fixed $\nu$ partition function $Z_{\nu}$ is dominated by the
$\theta$ vacuum with smaller energy density $e_0(\theta)$. From Eq.~(\ref{theta-expansion}),
it is clear that $e_0(\theta)$ has the lowest energy density at $\theta=0$. Thus, at
large $V$ and keeping $|\nu|/\sqrt{V}$ fixed, one employs a saddle point expansion around $\theta=0$
to obtain
\begin{eqnarray}   \label{Z-nu-saddle}
Z_{\nu}     &=& \frac{1}{2\pi V\chi_t}\,e^{-\frac{\nu^2}{2V\chi_t}}\lbrack
1- \frac{3\gamma}{V\chi_t^2} + \frac{6\gamma \nu^2}{V^2 \chi_t^3} - \frac{\gamma \nu^4}
{V^3 \chi_t^4}\rbrack.
\end{eqnarray}
 for a small $\gamma$.

   It has been shown that physical quantities in the $\theta$ vacuum, such as hadron masses,
can be extracted from fixed topological sectors~\cite{bcn03}. Consider the n-point Green's
function with operators $\mathcal{O}_i$ in the $\theta$ vacuum
\begin{eqnarray}   \label{G-theta}
G(\theta) &=& \langle \mathcal{O}_1\mathcal{O}_2...\mathcal{O}_n\rangle(\theta) \nonumber \\
         &=& \frac{1}{Z(\theta)} \int  \mathcal{D}A\mathcal{D}\overline{\Psi}\mathcal{D}\Psi
\mathcal{O}_1\mathcal{O}_2...\mathcal{O}_n e^{-S[A,\overline{\Psi},\Psi]} e^{i\theta Q[A]}.
\end{eqnarray}
  The corresponding Green's function in the topological sector $\nu$ is
\begin{equation}   \label{G-nu}
G_{\nu} =  \langle \mathcal{O}_1\mathcal{O}_2...\mathcal{O}_n\rangle_{\nu}
        = \frac{1}{Z_{\nu}} \frac{1}{2 \pi} \int_{-\pi}^{\pi} d\theta Z(\theta)
           G(\theta)\,e^{-i \theta \nu}.
\end{equation}

   Using the same saddle point approximation that was applied to $Z_{\nu}$ in
Eq.~(\ref{Z-nu-saddle}) to $G_{\nu}$, Eq.~(\ref{G-nu}) becomes
\begin{equation}
G_{\nu} = G(\theta_s) - \frac{12\gamma}{V\chi_t^2}\theta_s G'(\theta_s) +
\frac{1}{2V\chi_t}G''(\theta_s) (1 - \frac{12\gamma}{V\chi_t^2})
+ \frac{1}{8V^2\chi^2}G''''(\theta_s) + \mathcal{O}(\frac{1}{V^3}),
\end{equation}
where the pure imaginary $\theta$ at the saddle point is
\begin{equation}
\theta_s= -i \frac{\nu}{\langle Q^2\rangle}= -i \frac{\nu}{V\chi_t}.
\end{equation}

   This has been applied to two-point correlation function to show that
the mass which has the form $M(\theta) = M(0) + \frac{1}{2} M''(0)\theta^2$ in the $\theta$ vacuum
is calculated in topological sector $\nu$ is~\cite{bcn03} to be
\begin{equation}  \label{M-nu}
M_{\nu} = M(0) + \frac{1}{2} \frac{M''(0)}{V\chi_t}\lbrack 1- \frac{\nu^2}{V\chi_t}
- \gamma( \frac{12}{V\chi_t^2} -\frac{39 \nu^2}{V^2 \chi_t^3} + \frac{14\nu^4}{V^3 \chi_t^4}
- \frac{\nu^6}{V^4 \chi_t^5})\rbrack.
\end{equation}
Here, we have taken the full volume dependence into account.
From this expression, we see that the mass of the $\theta$-vacuum $M(0)$ can be
obtained from $M_{\nu}$ at several volumes or with several topological sectors,
or the combination thereof. Similarly, the topological susceptibility can be extracted
from Eq. (\ref{M-nu}) with several $\nu$ sectors. It is easy to prove that, upon summing over the
topological sectors with the probability $Z_{\nu}$ in Eq.~(\ref{Z-nu-saddle}), the
mass in the $\theta=0$ vacuum is recovered, i.e.
\begin{equation}   \label{M}
\frac{1}{\sum_{\nu} Z_{\nu}}\sum_{\nu} Z_{\nu} M_{\nu} = M(0).
\end{equation}
   $M_{\nu}$ in Eq.~(\ref{M-nu}) shows that the physical quantity measured
in the fixed topological sector is the same as that in the $\theta=0$ vacuum
with finite volume corrections. When the volume approaches infinity, the difference
goes away, so does the difference when results from different topological sectors
are added up and weighted by the probability of occurrence of the topological sector.

   In the present work, we extend this analysis to three-point functions with the
electromagnetic form factor in order to extract the  neutron electric dipole moment and
the anapole moment. For the case that $|\theta_s|$ is small, we can expand
$G_{\nu}$ around $\theta=0$ and obtain
\begin{eqnarray}
G_{\nu} &=& G(0) + (\frac{-i\nu}{2V\chi_t})G'(0)(1 - \frac{12\gamma}{V\chi_t^2})
+ \frac{G''(0)}{2V\chi_t}\lbrack
1 - \frac{\nu^2}{V\chi_t} - \frac{12 \gamma}{V\chi_t^2}\rbrack \nonumber \\
&+&\frac{-i\nu G'''(0)}{2 (V\chi_t)^2} + \frac{G''''(0)}{8V^2\chi_t^2} + \mathcal{O}(\frac{1}{V^3}).
\end{eqnarray}
When the Green's function is CP-even(odd), $G(\theta)$ is even(odd) in $\theta$,
the above expression holds separately for the even-odd cases as
\begin{eqnarray}
G_{\nu}^{even} &=&  G(0) + \frac{G''(0)}{2V\chi_t}\lbrack
1 - \frac{\nu^2}{V\chi_t} - \frac{12 \gamma}{V\chi_t^2} \rbrack + \frac{G''''(0)}{8V^2\chi_t^2}
+ \mathcal{O}(\frac{1}{V^3}), \label{G-nu-even}\\
G_{\nu}^{odd} &=& \frac{-i\nu}{V\chi_t}G'(0) (1 - \frac{12\gamma}{V\chi_t^2})
+\frac{-i\nu G'''(0)}{2 (V\chi_t)^2} + \mathcal{O}(\frac{1}{V^3}).  \label{G-nu-odd}
\end{eqnarray}

   It is now easy to see why the expression for evaluating the CP-odd quantities
that was alluded to in the introduction has the correct
large $V$ behavior. To evaluate CP-odd quantity at small $\theta$, it is sufficient
to consider the derivative of the corresponding path-integral representation of the
Green's function in Eq.~(\ref{G-theta}) w.r.t. $\theta$, i.e.
\begin{equation}  \label{G'}
G'(0)= i\langle Q \mathcal{O}\rangle = \frac{i}{\sum_{\nu} Z_{\nu}} \sum_{\nu} Z_{\nu} \nu \langle
\mathcal{O}\rangle_{\nu}.
\end{equation}
Plugging $Z_{\nu}$ from Eq.~(\ref{Z-nu-saddle}) and $\langle \mathcal{O}\rangle_{\nu}$
from Eq.~(\ref{G-nu-odd}) into Eq.~(\ref{G'}), one recovers $G'(0)$ to order $\frac{\gamma}{V}$.
It is straight-forward to show that had all the $V$ terms in Eq.~(\ref{G-nu-odd}) been kept, one
would have recovered $G'(0)$ exactly as is for $M(0)$ in Eq.~(\ref{M}).
Since the leading $1/V$ contribution of $\langle \mathcal{O}\rangle_{\nu}$  is $\frac{-i\nu}{2V\chi_t}G'(0)$
in Eq.~(\ref{G-nu-odd}), its contribution to $G'(0)$ in Eq.~(\ref{G'}) is $\propto \sum_{\nu} Z_{\nu} \frac{\nu^2}{V}$
which is independent of $V$ as it should. In other words, the apparent $\sqrt{V}$ dependence from
the explicit $\nu$ in Eq.~(\ref{G'}) does not lead to $\sqrt{V}$ divergence. Similarly,
it is easy to check that the $V\chi_t$ and $\nu$ dependence associated with $G''(0)$ and
$G'''(0)$ in Eqs.~(\ref{G-nu-even}) and (\ref{G-nu-odd}) are correct to reproduce the
the second and third derivatives of the path-integral formulation of $G(\theta)$.

\section{Neutron Electric Dipole Moment}

We want to extend the study of two-point function to three-point function with the electromagnetic form factors in the nucleon in order to extract the neutron eclectic dipole moment and anapole moment. The form factor for
the electromagnetic current $J_{\mu}^{EM}=i\sum_q\overline{\psi}_qe_q\gamma_{\mu}\psi_q$ in the nucleon is defined as
\begin{equation}
\langle p',s'|J_{\mu}^{EM}|p,s\rangle = \overline{u}_{s'}(p')\Gamma_{\mu}(q^2)u_s(p),
\end{equation}
where
\begin{equation}
\Gamma_{\mu}(q^2)=i\gamma_{\mu}F_1(q^2)-i\sigma_{\mu\nu}\frac{q_{\nu}}{2m_N}F_2(q^2)
+ (i\gamma_{\mu}\gamma_5q^2 -2m_N\gamma_5q_{\mu})F_A(q^2) -\sigma_{\mu\nu}\gamma_5
q_{\nu}\frac{F_3(q^2)}{2m_N},
\end{equation}
with $F_3(q^2)/2m_N$ and $F_A$ being the electric dipole and anapole form factors and
$\sigma_{\mu\nu}=\frac{1}{2i}[\gamma_{\mu},\gamma_{\nu}]$.
The electric dipole moment is
\begin{equation}
d_N= \frac{F_3(0)}{2m_N}.
\end{equation}
Note that due to the charge conservation
of the electromagnetic current $\partial_{\mu}J_{\mu}^{EM}=0$, one has the relation
\begin{equation}
\langle p',s'|\partial_{\mu}J_{\mu}^{EM}|p,s\rangle = \overline{u}_{s'}(p')q_{\mu}\Gamma_{\mu}(q^2)u_s(p) =0.
\end{equation}

The electromagnetic form factors, the neutron electric dipole and anapole form factors of the nucleon can be
obtained from the three-point Green's function~\cite{wdl92,sai05}. We shall extend it to the case at fixed topology.
Consider the three-point Green's function in a $\theta$ vacuum
\begin{equation} \label{path-integral}
G_{\theta}^{NJN}(\vec{q},t,t_f) = \langle \chi_N(p',t_f)J_{\mu}^{EM}(\vec{q},t)
\overline{\chi}_N(p,0)\rangle_{\theta},
\end{equation}
where $q= p'-p$ and $\chi_N$ is the nucleon interpolation field.
At large time separation, i.e. $t_f-t \gg 1$ and $t \gg 1$, the Green's function is
dominated by the lowest state which is the nucleon
\begin{eqnarray}   \label{three-pointA}
G_{NJN}(\theta,\vec{q},t,t_f)\!\!\!\!&&{}_{\stackrel{\longrightarrow}{t_f-t \gg 1,t \gg 1}}
 e^{-E_{N^{\theta}}' (t_f -t)}e^{-E_{N^{\theta}}t} \nonumber \\
& \times &\sum_{ss'} \langle \theta|\chi_N|N_{\theta}
(\vec{p'},s')\rangle \langle N_{\theta} (\vec{p'},s')|J_{\mu}^{EM}|N_{\theta} (\vec{p},s)\rangle
\langle N_{\theta} (\vec{p},s)|\overline{\chi}_N|\theta\rangle
\end{eqnarray}
where $|\theta\rangle = e^{i\theta Q}|0\rangle$ and $E_{N^{\theta}}'=\sqrt{\vec{p'}^2 +
m_{N^{\theta}}^2}, E_{N^{\theta}}=\sqrt{\vec{p}^2 + m_{N^{\theta}}^2}$.
The matrix elements for the interpolation fields are
\begin{eqnarray}
\langle \theta|\chi_N|N_{\theta}(\vec{p},s)\rangle = Z_N^{\theta}u_N^{\theta}(\vec{p},s)
\nonumber \\
\langle N_{\theta}(\vec{p'},s)|\overline{\chi}_N|\theta\rangle = Z_N^{\theta\,{}^*}\bar{u}_N^{\theta}(\vec{p'},s),
\end{eqnarray}
where the nucleon spinor projection is
\begin{equation}
\sum_s u_N^{\theta}(\vec{p},s)\bar{u}_N^{\theta}(\vec{p},s)=\frac{-i\gamma \cdot p + m_{N^{\theta}}
e^{i\alpha_N(\theta)\gamma_5}}{2E_{N^{\theta}}}.
\end{equation}
and the normalization is
\begin{equation}
\bar{u}_N^{\theta}(\vec{p},s')u_N^{\theta}(\vec{p},s)= \frac{m_{N^{\theta}}\,\cos\alpha_N(\theta)}
{E_N^{\theta}},
\end{equation}
The nucleon spinor satisfies the Dirac equation with a phase factor associated
with the mass term due to the CP-violation in the $\theta$ vacuum~\cite{sai05,sai07}
\begin{equation}
(i\gamma \cdot p + m_N^{\theta}e^{-i\alpha_N(\theta)\gamma_5}) u_N^{\theta}(\vec{p},s)
=\bar{u}_N^{\theta}(\vec{p},s)(i\gamma \cdot p + m_{N^{\theta}}e^{-i\alpha_N(\theta)\gamma_5})=0.
\end{equation}
Now the nucleon matrix element can be written as
\begin{equation}
\langle N_{\theta} (\vec{p'},s')|J_{\mu}^{EM}|N_{\theta} (\vec{p},s)\rangle
=\bar{u}_N^{\theta}(\vec{p'},s')\Gamma_{\mu}^{\theta}u_N^{\theta}(\vec{p},s),
\end{equation}
where the vertex $\Gamma_{\mu}^{\theta}$ can be separated in terms of CP-even and
CP-odd form factors
\begin{equation}
\Gamma_{\mu}^{\theta}=\Gamma_{\mu}^{even}(q,\theta)+\Gamma_{\mu}^{odd}(q,\theta),
\end{equation}

Since $\theta$ in QCD is small, one can consider small $\theta$ expansion
\begin{eqnarray}
\Gamma_{\mu}^{even}(q,\theta)\!\!\!&=&\!\!\! i\gamma_{\mu}F_1(q^2)-i\sigma_{\mu\nu}q_{\nu}F_2(q^2)
            +\frac{\theta^2}{2} \Gamma_{\mu}^{even\,''}(q,0)+... \nonumber \\
\Gamma_{\mu}^{odd}(q,\theta)\!\!\!&=&\!\!\! -\sigma_{\mu\nu}\gamma_5 q_{\nu}\frac{\theta F_3^{'}
(q^2,0)}{2m_N} + \theta F_A^{'}(q^2, 0)(i\gamma_{\mu}q^2 -2m_Nq_{\mu})\gamma_5
            + \frac{\theta^3}{3!} \Gamma_{\mu}^{odd\,'''}(q,0)+...
\end{eqnarray}
We have used the property that, for small $\theta$
\begin{eqnarray}
m_N^{\theta}& =& m_N + \frac{1}{2}m_N^{''}(0) \theta^2 + ... \nonumber \\
|Z_N^{\theta}|^2 &=& |Z_N|^2 + \frac{1}{2} |Z_N|^{2''}(0) \theta^2 + ... \nonumber \\
F_{1,2}(q^2,\theta)&=& F_{1,2}(q^2) + \frac{1}{2}F_{1,2}^{''}(q^2)\theta^2+...  \nonumber \\
F_{3,A}(q^2,\theta)&=& F_{3,A}^{'}(q^2,0)\theta + \frac{1}{3!}F_{3,A}^{'''}(q^2,0)\theta^3+...
\nonumber \\
\end{eqnarray}

     The three-point function in Eq.~(\ref{three-pointA}) is then
\begin{eqnarray}
G_{\theta}^{NJN}(\vec{q},t,t_f)_{\stackrel{\longrightarrow}{t_f-t \gg 1,t \gg 1}}\!\!\!\!\!\!\!
\!&& |Z_N|^2 e^{-E_{N^{\theta}}' (t_f -t)}e^{-E_{N^{\theta}}t} \frac{-i \gamma \cdot p' +
m_N^{\theta}e^{i\alpha_N(\theta)\gamma_5}}{2E_N^{'\theta}} \nonumber \\
&&\lbrack\Gamma_{\mu}^{even}(q,\theta)+\Gamma_{\mu}^{odd}(q,\theta)\rbrack
\frac{-i \gamma \cdot p + m_N^{\theta}e^{i\alpha_N(\theta)\gamma_5}}{2E_N^{\theta}}.
\end{eqnarray}
Expanding in $\theta$, it gives
\begin{eqnarray} \label{expansion}
&&G_{NJN}(\theta,\vec{q},t,t_f)_{\stackrel{\longrightarrow}{t_f-t \gg 1,t \gg 1}}\!\!\!\!\!\!\!
 |Z_N|^2 e^{-E_{N^{\theta}}' (t_f -t)}e^{-E_{N}t} \{\frac{-i \gamma \cdot p' +
m_N}{2E_N^{'}}\Gamma_{\mu}^{even}(q,0) \frac{-i \gamma \cdot p + m_N}{2E_N} \nonumber \\
&+& \theta \lbrack \frac{-i \gamma \cdot p' + m_N}{2E_N'}
\Gamma_{\mu}^{'odd}(q,0)\frac{-i \gamma \cdot p + m_N}{2E_N} \nonumber \\
&+& \frac{i\alpha_N' m_N\gamma_5}{2E_N'}\Gamma_{\mu}^{even}(q^2,0)\frac{-i \gamma \cdot p + m_N}{2E_N}
+ \frac{-i \gamma \cdot p' + m_N}{2E_N^{'}}\Gamma_{\mu}^{even}(q^2,0)
\frac{i\alpha_N' m_N\gamma_5}{2E_N}\rbrack \nonumber \\
&+& \frac{\theta^2}{2!} G_{NJN}^{''}(q, t_f - t \gg1, t \gg 1) +
\frac{\theta^3}{3!} G_{NJN}^{'''}(q, t_f - t \gg1, t \gg 1)+...\}
\end{eqnarray}

From the current experimental bound on the neutron electric dipole moment,
$\theta < 10^{-9}$ in QCD. Thus, it is sufficient to consider the CP-odd quantities
linear in $\theta$. Equating the linear $\theta$ term in Eq.~(\ref{expansion}) with that
in the $\theta$ expansion of the path-integral in Eq.~(\ref{path-integral})
\begin{equation} \label{theta_expansion}
G_{\theta}^{NJN}(\vec{q},t,t_f) = \langle \chi_N(p',t_f)J_{\mu}^{EM}(\vec{q},t)
\overline{\chi}_N(p,0)\rangle_{\theta} \approx \langle \chi_N(p',t_f)(1+i\theta Q)J_{\mu}^{EM}(\vec{q},t) \overline{\chi}_N(p,0)\rangle_{\theta=0},
\end{equation}
we obtain
\begin{equation}
G_{NJN}(\theta=0,\vec{q},t,t_f-t \gg 1) =
 |Z_N|^2 e^{-E_{N}' (t_f -t)}e^{-E_{N}t} \{\frac{-i \gamma \cdot p' +
m_N}{2E_N^{'}}\,\Gamma_{\mu}^{even}(q,0) \frac{-i \gamma \cdot p + m_N}{2E_N}\},
\end{equation}
and
\begin{eqnarray} \label{derivative}
G'_{NJN}(\theta=0,\vec{q},t,t_f-t \gg 1) &=& |Z_N|^2 e^{-E_{N'} (t_f -t)}e^{-E_{N}t}
\{\frac{-i \gamma \cdot p' +
m_N}{2E_N^{'}}\Gamma_{\mu}^{'\,odd}(q,0) \frac{-i \gamma \cdot p + m_N}{2E_N} \nonumber \\
&+& \frac{i\alpha_N'(0) m_N\gamma_5}{2E_N'}\Gamma_{\mu}^{even}(q^2,0)\frac{-i \gamma \cdot p + m_N}{2E_N}
\nonumber \\
&+& \frac{-i \gamma \cdot p' + m_N}{2E_N^{'}}\Gamma_{\mu}^{even}(q^2,0)
\frac{i\alpha_N'(0) m_N\gamma_5}{2E_N} \}
\end{eqnarray}
    $\alpha'(0)$ can be extracted from the nucleon correlator in the $\theta$ vacuum.
Consider
\begin{equation}
G_{NN}(\theta, \vec{p}, t)\equiv \langle \chi_N(\vec{p},t)\overline{\chi}_N(\vec{p},0)e^{i\theta Q}\rangle_{\theta}
= G_{NN}(0, \vec{p}, t) + \theta  G_{NN}'(0, \vec{p}, t) + O(\theta^2) + ...
\end{equation}
Taking the appropriate trace and asymptotic limit of the correlator with $t \gg 1$, we obtain
\begin{eqnarray} \label{two-point}
Tr  \langle \Gamma_4 \chi_N(\vec{p},t)\overline{\chi}_N(\vec{p},0)\rangle_{\theta=0}
\equiv Tr[\Gamma_4 G_{NN}(0,\vec{p},t)_{\stackrel {\longrightarrow}{t \gg 1}}
|Z|^2 \frac{E_N + m}{E_N}e^{-E_N t} \nonumber \\
Tr \langle \gamma_5 Q \chi_N(\vec{p},t)\overline{\chi}_N(\vec{p},0)\rangle_{\theta=0} \equiv
-iTr[\gamma_5 G_{NN}(0,\vec{p},t)_{\stackrel{\longrightarrow}{t \gg 1}} |Z|^2 \frac{2 m}{E_N}\alpha'(0)e^{-E_N t}
\end{eqnarray}
where $\Gamma_4 = \frac{1+\gamma_4}{2}$ is the projection operator for the time-forward nucleon. If the
time $t$ is not large enough to filter out the negative parity excited nucleon $S_{11}$, one will need
to use the projector $1+ \frac{m_{-}}{E_{-}}\gamma_4$ where $m_{-}/E_{-}$ is the mass/energy of $S_{11}$.

From the ratio of the two-point functions, one can obtain $\alpha'(0)$, i.e.
\begin{equation}
\frac{E_N+m}{2m} \frac{Tr \langle \gamma_5 Q \chi_N(\vec{p},t)\overline{\chi}_N(\vec{p},0)\rangle}
{Tr  \langle \Gamma_4 \chi_N(\vec{p},t)\overline{\chi}_N(\vec{p},0)\rangle}\,\,\,{}_{\stackrel{-\!\longrightarrow}{t \gg 1}}
\,\,\, \alpha'(0).
\end{equation}

  In the case of fixed topology that we consider in this manuscript, one can consider
the two-point functions $G_{NN}^{\nu}$ in a specific topological sector $\nu$
\begin{equation} \label{two-point-nu}
Tr [\Gamma_4 G_{NN}^{\nu}(\vec{p},t)]_{\stackrel{\longrightarrow}{t \gg 1}}
|Z|^2 e^{-E_N t}\frac{E_N + m_N}{E_N} + \frac{1 -\nu^2/V\chi_t - 12\gamma/V\chi_t^2}{2V\chi_t}
Tr [\Gamma_4 G_{NN}^{''}(0,\vec{p}, t \gg 1) + ...
\end{equation}
and
\begin{equation}  \label{two-point-gamma5}
Tr [\gamma_5 G_{NN}^{\nu}(\vec{p},t)]_{\stackrel{\longrightarrow}{t \gg 1}}
|Z|^2 e^{-E_N t}\frac{\nu (1- 12\gamma/V\chi_t^2)}{V\chi_t}\frac{m_N}{E_N} \alpha'(0)
+\frac{-i \nu}{2!V^2\chi_t^2} Tr[\gamma_5 G_{NN}^{'''}(0, \vec{p}, t \gg 1)]
+ ...
\end{equation}
With several volumes, one can fit Eqs.~(\ref{two-point-nu}) and (\ref{two-point-gamma5}) and obtain
$\alpha'(0), |Z|^2, m_N$, and $E_N$.

   As for the form factors, one can consider the following three-point functions at a fixed
topology with the sink nucleon momentum $\vec{p}' = 0$:
\begin{eqnarray}  \label{3-44}
Tr [\Gamma_4 G_{NJ_4N}^{\nu}(\vec{p},t,t_f)]_{\stackrel{\longrightarrow}{t,t_f-t \gg 1}}
|Z|^2 e^{-m_N(t_f-t)}e^{-E_N t}\frac{E_N + m_N}{E_N}[iG_E(q^2)] \nonumber \\
+ \frac{1 -\nu^2/V\chi_t - 12\gamma/V\chi_t^2}{2V\chi_t}
Tr [\Gamma_4 G_{Nj_4N}^{''}(0,\vec{p}, t_f,t \gg 1)] + ...
\end{eqnarray}
where $G_E(q^2) = F_1(q^2) + \frac{q^2}{4m_N^2}F_2(q^2)$ is the electric form factor
with $q^2 = (E_N -m_N)^2 - \vec{q}^2$ where $\vec{q} = - \vec{p}$.
\bigskip
\begin{eqnarray}  \label{3-45ji}
&&\,\,\,\,\,Tr[i\Gamma_4 \gamma_5\gamma_jG_{NJ_iN}^{\nu}(\vec{p},t,t_f)]
_{\stackrel{\longrightarrow}{t,t_f-t \gg 1}}
|Z|^2 e^{-m_N(t_f-t)}e^{-E_N t} \nonumber \\
&&\times \Big{\{} i\epsilon_{ijk}G_M(q^2)
+ \frac{-i\nu(1-12\gamma/V\chi_t^2)}{V\chi_t} \frac{iq_iq_j}{4m_NE_N}[F_3'(q^2,0) -i 4m_N^2 F_A'(q^2,0)
+2m_N F_2(q^2)\alpha'(0)]\Big{\}}  \nonumber \\
&&+ \,\,\,\,\frac{1 -\nu^2-12\gamma/V\chi_t^2}{2V\chi_t}
Tr [i\Gamma_4 \gamma_5\gamma_jG_{NJ_iN}{''}(0,\vec{p}, t_f,t \gg 1)] + ...
\end{eqnarray}
where $G_M(q^2) = F_1(q^2) + F_2(q^2)$ is the magnetic form factor.
\bigskip
\begin{eqnarray}  \label{3-544}
&&Tr[\gamma_5\Gamma_4 G_{NJ_4N}^{\nu}(\vec{p},t,t_f)]
_{\stackrel{\longrightarrow}{t,t_f-t \gg 1}}
|Z|^2 e^{-m_N(t_f-t)}e^{-E_N t} \nonumber \\
&&\times \frac{i\nu(1-12\gamma/V\chi_t^2)}{V\chi_t}\Big{\{} \frac{\vec{q}^2}{2E_Nm_N} F_3'(q^2,0)
+ [\frac{E_N+m_N}{2E_N} F_1(q^2)
+ \frac{\vec{q}^2}{4m_NE_N}F_2(q^2)]\alpha'(0)\Big{\}}  \nonumber \\
&&+ \,\,\,\,\frac{1 -\nu^2/V\chi_t-12\gamma/V\chi_t^2}{2V\chi_t} Tr [\gamma_5\Gamma_4G_{NJ_4N}{''}(0,\vec{p},
t_f,t \gg 1)]
+ ...
\end{eqnarray}
and
\begin{eqnarray}   \label{3-45i4}
&&Tr[i\Gamma_4 \gamma_5\gamma_iG_{NJ_4N}^{\nu}(\vec{p},t,t_f)]
_{\stackrel{\longrightarrow}{t,t_f-t \gg 1}}
|Z|^2 e^{-m_N(t_f-t)}e^{-E_N t} \nonumber \\
&&\times \frac{-i\nu(1-12\gamma/V\chi_t^2)}{V\chi_t}\frac{q_i}{2E_N}\Big{\{} \frac{E_N+m_N}{m_N} F_3'(q^2,0)
+ [F_1(q^2) + \frac{E_N+3m_N}{2m_N} F_2(q^2)]\alpha'(0)\Big{\}}  \nonumber \\
&&+ \,\,\,\,\frac{1 -\nu^2/V\chi_t-12\gamma/V\chi_t^2}{2V\chi_t} Tr [i\Gamma_4 \gamma_5\gamma_iG_{NJ_4N}''(0,\vec{p},
t_f,t \gg 1)]
+ ...
\end{eqnarray}

  Combining the two-point functions in Eqs.~(\ref{two-point-nu}) and (\ref{two-point-gamma5})
and three-point functions in Eqs.~(\ref{3-44}), (\ref{3-45ji}), (\ref{3-544}), and (\ref{3-45i4}) for
several volumes, one can extract the neutron electric dipole form factor
\begin{equation}
d_N(q^2)=\theta F_3'(0,q^2),
\end{equation}
the anapole form factor
\begin{equation}
a_N(q^2)=\theta F_A'(0,q^2),
\end{equation}
in addition to the electric and magnetic form factors
$G_E(q^2)$ and $G_M(q^2)$ (and/or  $F_1(q^2)$ and $F_2(q^2)$). When and if the experimental results on the dipole and anapole moments are known to be due to the QCD $\theta$ term, one can then determine $\theta$.

\section{Zero Modes, Quenched Approximation, and Anomalous Magnetic Form Factor}

   The role of zero modes in the quark propagators has been discussed extensively in the
literature. For example, in the quark condensate, its contribution is $\frac{2|\nu|}{m_qV}$ for
a configuration with topological charge $\nu$ in the $\theta$ vacuum~\cite{ehn99}. Similarly,
the pion correlator has a leading $\frac{1}{m_q^2V^2}$ contribution from the zero modes~\cite{blu00,ddh02}. Since
one should take the infinite volume limit before the chiral limit to have chiral symmetry
breaking manifested, the zero mode contribution goes away in these limits or their contributions become
negligible when the volume is sufficient large. However, for CP-odd quantities such as the neutron electric
dipole moment, the situation is different.  We see from Eqs.~(\ref{G-nu-odd}), (\ref{3-45ji}), (\ref{3-544}),
and (\ref{3-45i4}) that, for CP-odd Green's functions, the leading contribution is proportional to $\frac{1}{V}$.
Therefore, the zero modes will contribute at the finite volume. Specifically, for the case that only one
of the quark propagators involves zero modes, it would yield a $\frac{1}{V}$ factor and will, thus, contribute to
the CP-odd Green's functions to leading order in $\frac{1}{V}$.

   For the quenched approximation, the zero mode contribution to the neutron electric dipole
moment will lead to power divergence in $m_q$ at the chiral limit in a finite volume. This divergence is
not protected by the fermion determinant as is in the dynamical fermion case where the determinant measure
is proportional to $m_q^{|\nu|N_F}$ and will cancel out the mass singularity in the quark
propagators due to zero modes. In this case, the divergence is $\frac{1}{m_q^3}$
for $\nu=1$, and $\frac{1}{m_q^4}$ for $\nu \ge 2$ which are inherent in the quenched approximation.
 Therefore, it does not make sense to consider NEDM in the quenched approximation, except perhaps to test the
algorithm.

   It has been shown~\cite{bgh06} recently that, in the light-cone formalism, there is a universal
relation between the electric dipole form factor and the anomalous magnetic form factor for each Fock state, i.e.
\begin{equation}  \label{F3-F2}
[F_3(q^2)]_a = tan \beta_a [F_2(q^2)]_a
\end{equation}
where $a$ denotes the Fock state and $\beta_a$ is the $P_{\bot}-$ and $T_{\bot}$-violating phase in the
Fock state $a$. Since $\beta_a$ depends on $a$, $F_3(q^2)$ and $F_2(q^2)$, being the sum of their Fock state
components, do not have the same $q^2$ dependence in general, unless the Fock state sum is saturated by
a single Fock state or $\beta_a$ is a pure constant independent of $a$~\cite{gar08}.
We do not see a proportionality between $F_3'(q^2,0)$ and $F_2(q^2)$ in the present formulation. Even though the
CP-odd three-point functions in all $\nu \neq 0$
topological sectors must have the same $q^2$ dependence in order that Eq.~(\ref{G'}) is satisfied, $F_3'(q^2)$
and $F_2(q^2)$ can have different $q^2$ behaviors in the various projected
three-point functions considered in Eqs.~(\ref{3-44}), (\ref{3-45ji}), (\ref{3-544}), and (\ref{3-45i4}).
With precise enough lattice simulations, one should be able to check if $F_3(q^2)$ is proportional to $F_2(q^2)$.
One can also check the conjecture~\cite{bgh06} that the neutron and proton electric dipole moments repeat the
isospin structure of the anomalous magnetic moments, i.e. $|d^n +d^p| \ll |d^n - d^p|$.

\section{Conclusion}

   We have studied the Green's function at fixed topological sector for the CP-odd case and
clarified a question regarding the large $V$ behavior. As an application, we
have formulated the three-point functions for the electromagnetic current in the nucleon at fixed topology
which are needed to extract the neutron electric dipole and anapole form factors in this practical
calculations. It is shown that, in
the quenched approximation, the zero modes lead to power divergence in the quark mass at the chiral
limit for the CP-odd Green functions.

This work is supported by U.S. DOE grant DE-FG05-84ER40154.
The author wishes to thank S. Aoki, S. Brodsky, P. deForcrand, S. Gardner, P. Hasenfratz, and M. L\"{u}scher
for useful discussions. He also acknowledge the hospitality received while visiting
CERN where this work is initiated in Oct. 2006.



\begin{thebibliography}{99}



\bibitem{bbo06}
F. Berruto, T. Blum, K. Orginos, and A. Soni
Phys.Rev. {\bf D73}, 054509 (2006), [hep-lat/0512004].

\bibitem{sai05}
E. Shintani et al., Phys.Rev. {\bf D72}, 014504 (2005), [hep-lat/0505022].

\bibitem{sai07}
E. Shintani, S. Aoki, N. Ishizuka, K. Kanaya, Y. Kikukawa, Y. Kuramashi, M. Okawa,
A. Ukawa, and T. Yoshi\'{e}, Phys. Rev. {\bf D75}, 034507 (2007), [hep-lat/0611032].

\bibitem{sai08}
E. Shintani, S. Aoki, and Y. Kuramashi, [arXiv:0803.0797].

\bibitem{fhh06}
H.Fukaya, S.Hashimoto, T.Hirohashi, K.Ogawa, T.Onogi, Phys. Rev. {\bf D73}, 014503 (2006),
[hep-lat/0510116].

\bibitem{iva06}
I. Horv'{a}th, [hep-lat/0605008].

\bibitem{ahl08}
A. Alexandru, I. Horv'{a}th, and K.F. Liu, [arXiv:0803.2744].

\bibitem{liu06}
K.F. Liu, PoS {\bf LAT2006}, 056 (2006), [hep-lat/0609033].

\bibitem{streuer07}
T. Streuer, talk at Lattice 2007.

\bibitem{bcn03}
R. Brower, S. Chandrasekharan, J.W. Negeleb and U.-J. Wiese,
Phys. Lett. {\bf B560}, 64 (2003).

\bibitem{afh07}
S. Aoki, H. Fukaya, S. Hashimoto, and T. Onogi, Phys. Rev. {\bf D76}, 054508 (2007),
[arXiv:0707.0396].

\bibitem{ls92}
H. Leutwyler and A. Smiliga, Phys. Rev. {\bf 46}, 5607 (1992).

\bibitem{neu98}
H. Neuberger, Phys. Lett. {\bf B 417}, 141 (1998).

\bibitem{wit80}
E. Witten, Annals Phys. {\bf 128}, 363 (1980).

\bibitem{ly52}
T.D. Lee and C.N. Yang, Phys. Rev. {\bf 87}, 404, 410 (1952).

\bibitem{wdl92}
W. Wilcox, T. Draper, and K.F. Liu, Phys. Rev. {\bf D46}, 1109 (1992), [hep-lat/9205015].

\bibitem{ehn99}
R.G. Edwards, U.M. Heller, and R. Narayanan, Phys. Rev.{\bf D59}, 094510 (1999).

\bibitem{blu00}
T. Blum, et al., Phys. Rev. {\bf D69}, 074502 (2004); [arXiv:hep-lat/0007038].

\bibitem{ddh02}
S.J. Dong, T. Draper, I. Horvath, F.X. Lee, K.F. Liu, and J.B. Zhang,
Phys. Rev. {\bf D65}, 054507 (2002), [arXiv:hep-lat/0108020].

\bibitem{bgh06}
S. J. Brodsky, S.  Gardner, and D.S. Hwang, Phys. Rev. {\bf D73}, 036007 (2006).

\bibitem{gar08}
S. Gardner, private communication.

\end{thebibliography}
\end{document}